\documentclass{XrU2005}
\usepackage{graphicx}

\title{New insights into ultraluminous X-ray sources from deep {\it
XMM-Newton\/} observations}
\author{T.P. Roberts}
\author{A.-M. Stobbart}
\author{M.R. Goad}
\author{R.S.Warwick}
\affil{X-ray \& Observational Astronomy Group, Dept. of Physics \& Astronomy, University of Leicester, University Road, Leicester, LE1 7RH, UK}
\author{J. Wilms}
\affil{Astronomy \& Astrophysics Group, Department of Physics, University of Warwick, Coventry, CV8 1GA, UK}
\author[3]{P. Uttley}
\author[3,4]{J.N. Reeves}
\affil{Exploration of the Universe Division, NASA Goddard Space Flight Center, Greenbelt Road, Greenbelt, MD 20771, USA}
\affil{Dept. of Physics \& Astronomy, Johns Hopkins University, 3400 N Charles Street, Baltimore, MD 21218, USA}

\begin{document}

\keywords{black hole physics - X-rays: binaries - X-rays: galaxies}

\maketitle

\begin{abstract}
The controversy over whether ultraluminous X-ray sources (ULXs)
contain a new intermediate-mass class of black holes (IMBHs) remains
unresolved.  We present new analyses of the deepest {\it XMM-Newton\/}
observations of ULXs that address their underlying nature.  We examine
both empirical and physical modelling of the X-ray spectra of a sample
of thirteen of the highest quality ULX datasets, and find that there
are anomalies in modelling ULXs as accreting IMBHs with properties
simply scaled-up from Galactic black holes.  Most notably, spectral
curvature above 2 keV in several sources implies the presence of an
optically-thick, cool corona.  We also present a new analysis of a 100
ks observation of Holmberg II X-1, in which a rigorous analysis of the
temporal data limits the mass of its black hole to no more than 100
M$_{\odot}$.  We argue that a combination of these results points
towards many (though not necessarily all) ULXs containing black holes
that are at most a few 10s of M$_{\odot}$ in size.
\end{abstract}

\section{ULXs and IMBHs}

ULXs have garnered a great detail of attention since the launch of
{\it Chandra\/} and {\it XMM-Newton\/}, as these missions have
provided the first opportunity to study these remarkable objects in
great detail (see e.g. \citealt{MC04}).  The key question in these
sources is whether their extreme X-ray luminosities originate in
isotropic radiation from a compact object accreting below the
Eddington rate - requiring the presence of an IMBH - or whether it can
be explained by another means, with the prime suspects being
anisotropic or super-Eddington radiation from a stellar-mass black
hole.

The outstanding piece of recent evidence in support of the presence of
IMBHs in ULXs derives from fitting their X-ray spectra with the same
empirical model as used for Galactic black hole X-ray binaries
(BHXRBs).  It has been found that a number of luminous ULXs are well
fitted by the combination of a soft multi-colour disc blackbody
(MCDBB, representative of an accretion disc spectrum) plus a hard
power-law continuum model (representative of a hot, optically thin
corona).  However, there is one crucial difference: the temperature of
the accretion disc in these ULXs is a factor $\sim 10$ lower, at $\sim
0.1 - 0.2$ keV, than in Galactic systems.  As the temperature of the
innermost edge of an accretion disc scales with the mass of the
compact object as $kT_{\rm in} \propto M_{\rm BH}^{-0.25}$, this
implies very massive black holes in these ULXs, at $\sim 1000$
M$_{\odot}$ (e.g. \citealt{MFMF03}; \citealt{MFM04a}).

However, recent analyses of high spectral quality ULX data have shown
that some ULXs are best described by a variant of this empirical model
where the power-law continuum dominates {\it at low energies\/}
(e.g. \citealt{SRW04}; \citealt{Fos04}), and the disc component is
similar to that observed in stellar-mass black holes.  Furthermore, in
the case of NGC 5204 X-1 there is a clear ambiguity, with both model
variants fitting the same data (\citealt{RWWGJ05}).  So a {\it bona
fide\/} ``alternate'' empirical description for the spectra of some
ULXs does exist, albeit one with serious physical challenges.  In
particular the origin of the dominant soft power-law is unclear - it
appears too soft to originate in a jet, and cannot be
disc-Comptonisation as it dominates well below the peak emissivity of
the disc spectrum.

\section{A sample of bright ULXs}

The existence of this second class of ULX spectral shape, and the
ambiguity between this model and the ``IMBH'' spectrum in NGC 5204
X-1, led us to consider the following questions:

\begin{itemize}
\item
How easy is it to differentiate the two spectral forms?
\item
How common is each type of spectrum?
\item
What are the physics underlying the alternate spectral model?
\end{itemize}

We have therefore selected and uniformly reduced a sample of 13 ULXs,
comprising the highest quality EPIC spectral data available from the
{\it XMM-Newton\/} archive, to address these questions.  The data were
primarily selected on the basis of datasets with $\sim 10$ ks or more
EPIC exposure, and a measured {\it ROSAT\/} High Resolution Imager
count rate $> 10$ ct ks$^{-1}$ (taken from \citealt{RW2000} or
\citealt{CP02}).  Though this sample is small, the ULXs are
representative of the full range of ULX luminosity ($\sim 10^{39} - 2
\times 10^{40}$ erg s$^{-1}$), and with a minimum of a few thousand
counts per source represent the best defined X-ray spectra of ULXs to
date.  More details on this work will appear in \cite{SRW06}.

\subsection{Empirical spectral models}

Our initial spectral fits were made utilising simple single component
spectral models, subject only to absorption by material along the
line-of-sight to the ULXs.  The high definition and underlying
complexity of the ULX spectra were highlighted by these fits, with
none of the sources being well-fit (using a 95\% probability of
rejection criterion) by an absorbed MCDBB model, and only 5/13 being
adequately fit by a power-law continuum (including the three lowest
quality datasets).

The use of two-component models improved the fits considerably.  In
particular, the IMBH model (i.e. cool MCDBB + hard power-law) produced
good fits to 8/13 datasets (using the same criterion as above), with
an inner-disc temperature $kT_{\rm in} \sim 0.1 - 0.25$ keV and a
power-law photon index $\Gamma \sim 1.6 - 2.5$.  Again, the disc
temperatures inferred in these sources lead to black hole mass
estimates in the range $\sim 1000$ M$_{\odot}$.  However, the anomaly
in power-law slope discussed by \cite{RWWGJ05} remains - though one
might reasonably expect the IMBH accretion disc to be in a
``high/soft'' or ``very high'' state, given that it appears to be
accreting at or above roughly 10\% of the Eddington rate, these states
typically (though not exclusively) show power-law continua with photon
index $\Gamma > 2.4$ (\citealt{MR03}).  The power-law continua
exhibited by these IMBH candidates therefore appear to possess photon
indices that are somewhat on the low side.

The alternate empirical model (soft power-law, dominant at low
energies, plus a warm MCDBB) also produced a total of 8/13 good fits.
In fact, six of these sources were also well-fit by the IMBH model,
hence spectral ambiguity is present in 6/13 of the ULXs even in our
high data quality sample.  Of the remaining sources, two apiece were
uniquely well-fit by either the IMBH or alternate model, and the
remaining three were well-fit by neither (though two favoured an IMBH
fit, and one the alternate).

There is a potential discriminator between the two models that allows
us to investigate which model the ambiguous sources prefer, namely
curvature in the 2--10\,keV range.  This should be present in the
alternate model (which is dominated by the MCDBB in this regime) but
not the power-law-dominated IMBH model.  We therefore examined the
2--10\,keV data for each ULX by fitting both power-law and broken
power-law models, and looking for a significant improvement between
the two fits.  This approach was vindicated by demonstrating that
those sources best fit by the alternate model showed strong curvature
($> 4 \sigma$ improvements in the broken power-law fit over the
power-law fit, according to the F-test).  Of the ambiguous sources,
3/6 also showed some evidence for curvature ($> 2 \sigma$
significance) as, rather surprisingly, did two of the IMBH model
sources.  Hence we find at least marginal evidence for 2--10\,keV
curvature in $> 50\%$ of our spectra.

The above results suggests that the alternate model is at least a
viable description of the spectrum of more than half of our sample.
Hence we are once again faced with the problem of the dominant soft
power-law.  However, as there is no strong physical motivation for
using a power-law to describe the soft excess apparent in these
sources, we decided to test substituting another component, with the
constraint that it should have morphological similarities to an
absorbed power-law.  We therefore attempted fits using a classical
black body to describe the soft excess, and retained a MCDBB as the
harder component (we describe this as a ``dual thermal'' model).  We
found this to be the most successful empirical description of the
data, producing good fits to 10/13 ULXs, with typical parameters of
$kT_{\rm BB} \sim 0.25$ keV and $kT_{\rm in} \sim 1 - 2$ keV.  We
speculate that this could describe the spectrum of the accretion disc
around a stellar-mass black hole, with a related optically-thick
outflow (or ``black hole wind'', c.f. \citealt{KP03}) producing the
soft excess.

\begin{figure*}
\centering
\includegraphics[width=8cm,angle=270]{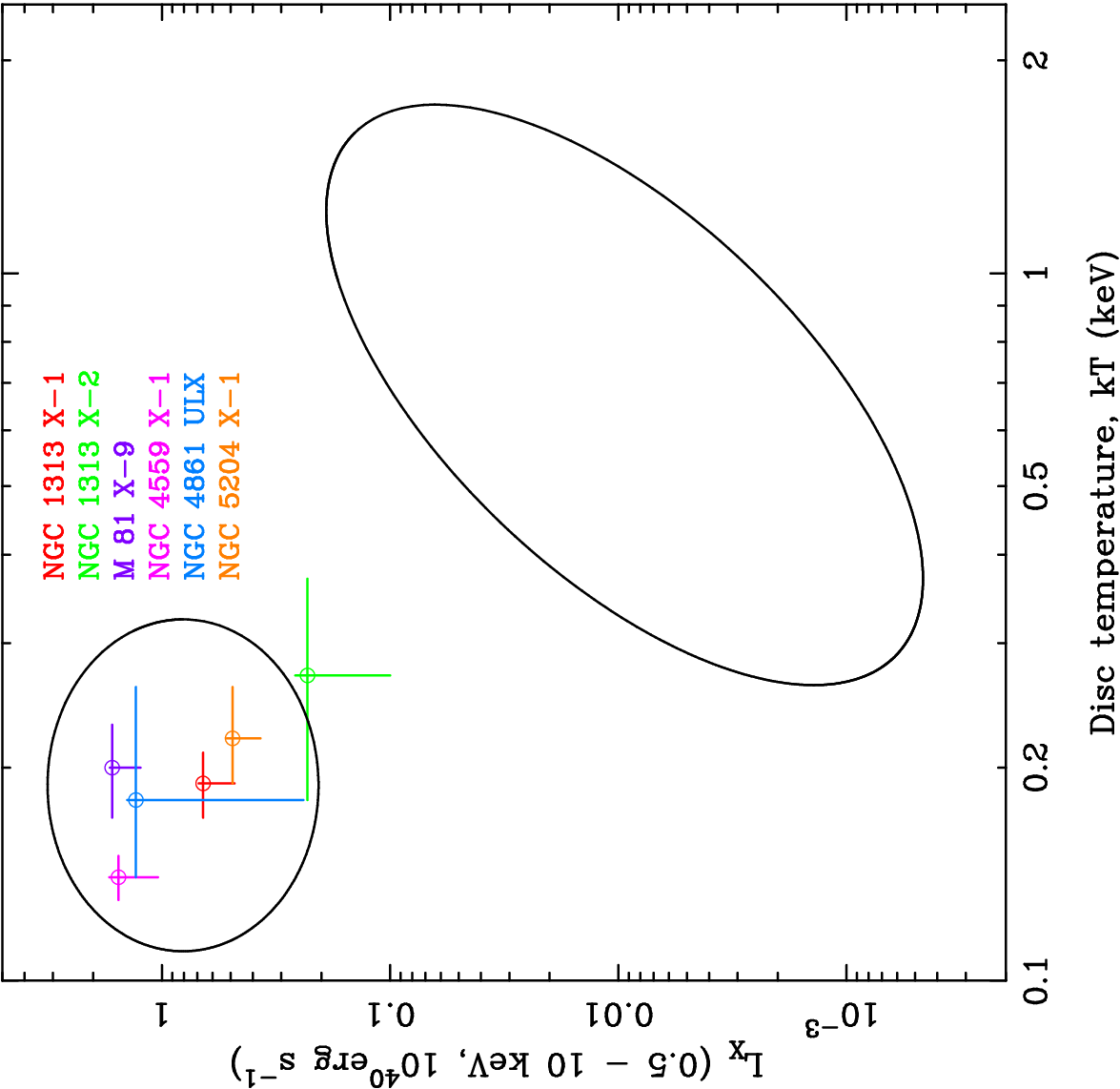}
\hspace*{0.5cm}
\includegraphics[width=8cm,angle=270]{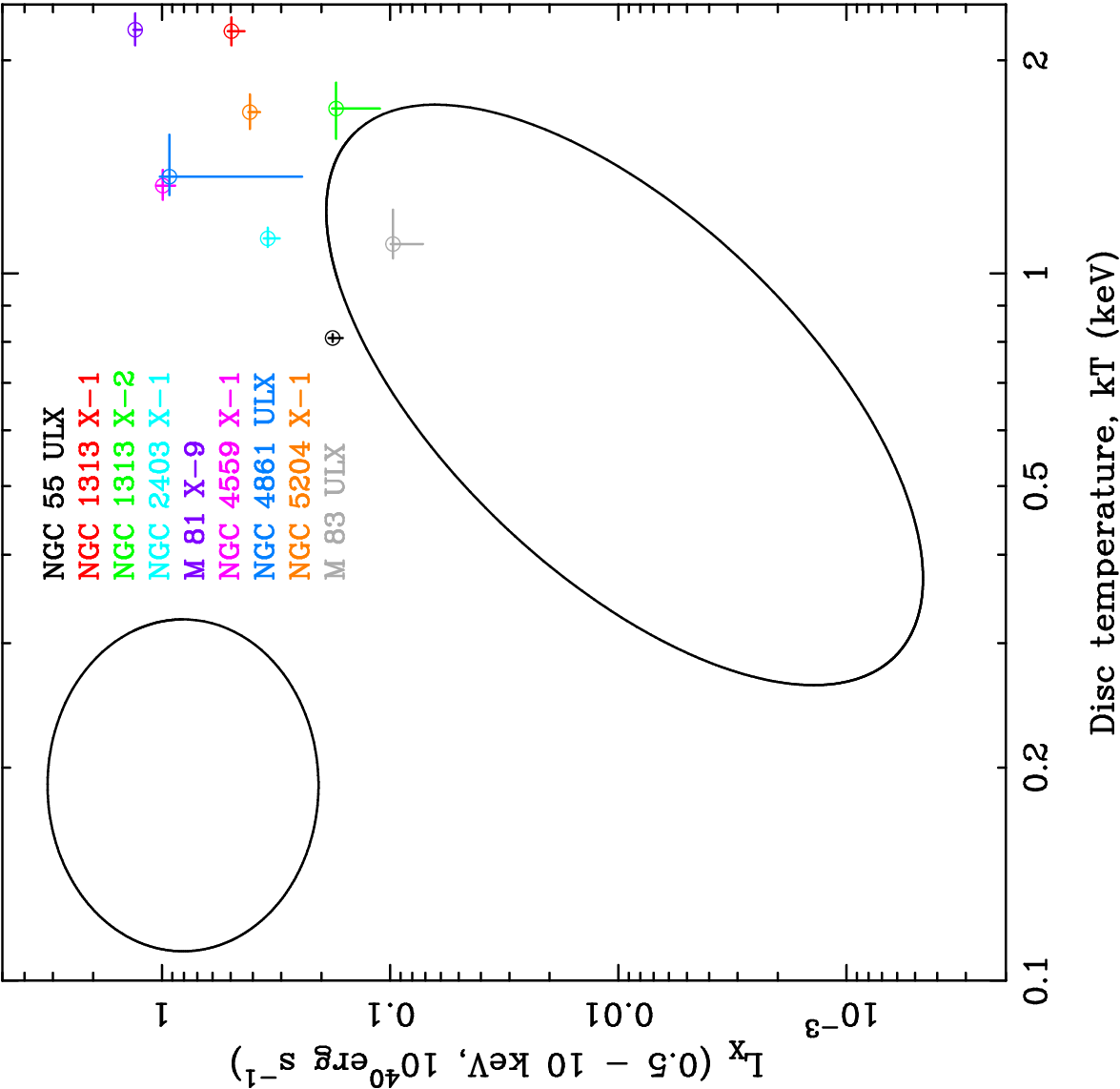}
\caption{The relationship between accretion disc luminosity and
temperature, calculated and displayed as per \cite{MFM04b}.  The
ellipses represent the regions in which the IMBH candidates (top left)
and stellar-mass black holes (bottom right) lay in Figs. 1 \& 2 of
that paper.  {\it Left panel\/}: We overlay data points from our IMBH
fits (note that we only show 6 of 8 points; of the missing pair, one
had a lower temperature [$\sim$ 80 eV] whilst the other had a
temperature that was unconstrained).  {\it Right panel\/}: The data
points from our dual thermal model overlayed (in this case one disc
was too hot to appear within the plotted range).}
\label{fig1}
\end{figure*}

We demonstrate one interesting implication of these fits in
Fig.~\ref{fig1}.  We essentially re-plot the results of
\cite{MFM04b} as elliptical regions in $kT_{\rm in} - L_{\rm X}$ space,
representing the high-luminosity, low temperature IMBH candidate discs
in the top left-hand corner, and the lower luminosity, warm discs of
stellar-mass black holes in our own Galaxy in the bottom-right.  In
the left panel we demonstrate that when we model our sample of ULXs as
IMBHs, we reproduce the results of \cite{MFM04b}, that is that the
IMBH candidates sit in a separate part of $kT_{\rm in} - L_{\rm X}$
parameter space from stellar-mass black holes.  However, the right
panel demonstrates that when we use the (more successful) dual-thermal
model, the ULXs appear to sit in a direct, high luminosity
continuation of the stellar-mass black hole relationship (broadly
$L_{\rm X} \propto T^4$ as the accretion rate increases, as expected
for standard accretion discs).  This evidence shows that, at the very
least, the mass of the underlying black hole inferred from empirical
spectral fitting of ULXs appears very dependent on the choice of
model.

\subsection{Physical spectral models}

We next investigated the origin of the possible 2--10\,keV curvature
using more physically-motivated models.  The ``slim disc'' model of an
accretion disc (e.g. \citealt{Wat01}; \textsc{xspec} parameterisation
courtesy K. Ebisawa) was only successful in fitting our spectra in
3/13 cases.  Instead, we found a physically self-consistent accretion
disc plus Comptonised corona model, using a {\tt diskpn + eqpair}
model in \textsc{xspec} (\citealt{GMNE01}, \citealt{Coppi00}), gave
good fits to 11/13 sources (plus a further source that was only
marginally rejected at the 95\% criterion).  All the fits display a
cool disc component, with temperatures $\sim 0.1 - 0.3$ keV, as would
be expected from IMBHs.  However, the majority of these fits also show
a second remarkable characteristic; the spectral curvature present in
many sources originates in a coronal component that is optically
thick, with typical optical depths of $\tau \sim 10 - 40$.  This is
very puzzling, because if ULXs are to be understood as direct
higher-mass analogies of high-state BHXRBs, then their coronae should
presumably be similarly optically thin ($\tau < 1$).  This only occurs
in two of our sources, leaving this pair as the best accreting IMBH
candidate ULXs within our sample.  For the remaining ULXs, that
possess apparent optically-thick coronae, it is evident that they are
not simple scaled-up Galactic BHXRBs in a classic high/soft state, and
we may perhaps need new insight to explain their spectra.

\section{Holmberg II X-1}

This source is regarded as the archetypal luminous ($L_{\rm X} >
10^{40}$ erg s$^{-1}$) nearby ULX, and has been widely studied both in
X-rays (e.g. \citealt{ejit04}) and over complementary wavebands
(e.g. \citealt{KWZ04}; \citealt{MMN05}).  We were awarded a 100-ks
observation of this source in \textit{XMM-Newton} AO-3, the results of
which we summarise here (see also \citealt{GRRU05}).

\begin{figure*}
\centering
\includegraphics[width=5.5cm,angle=270]{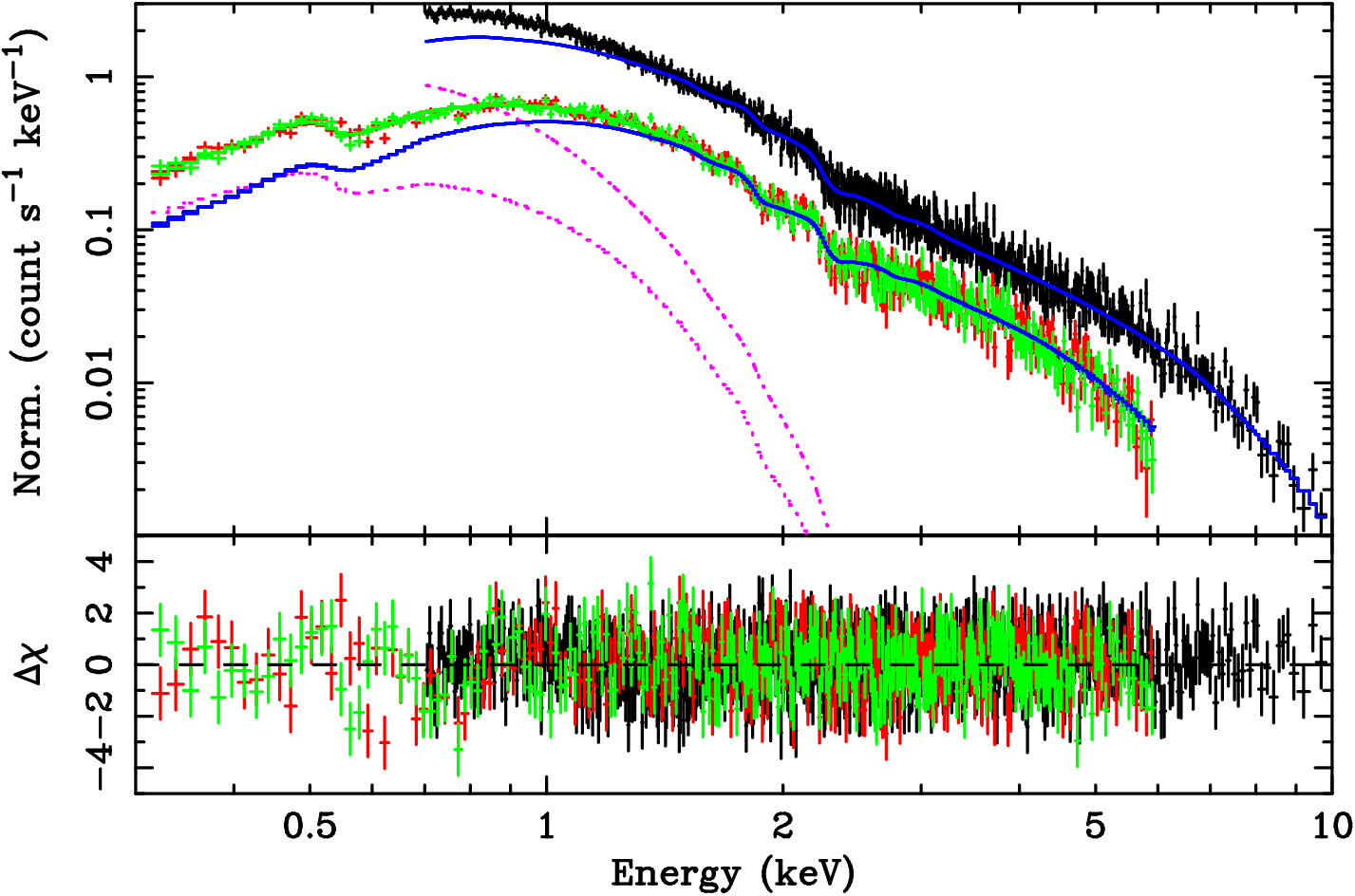}
\hspace*{0.5cm}
\includegraphics[width=5.5cm,angle=270]{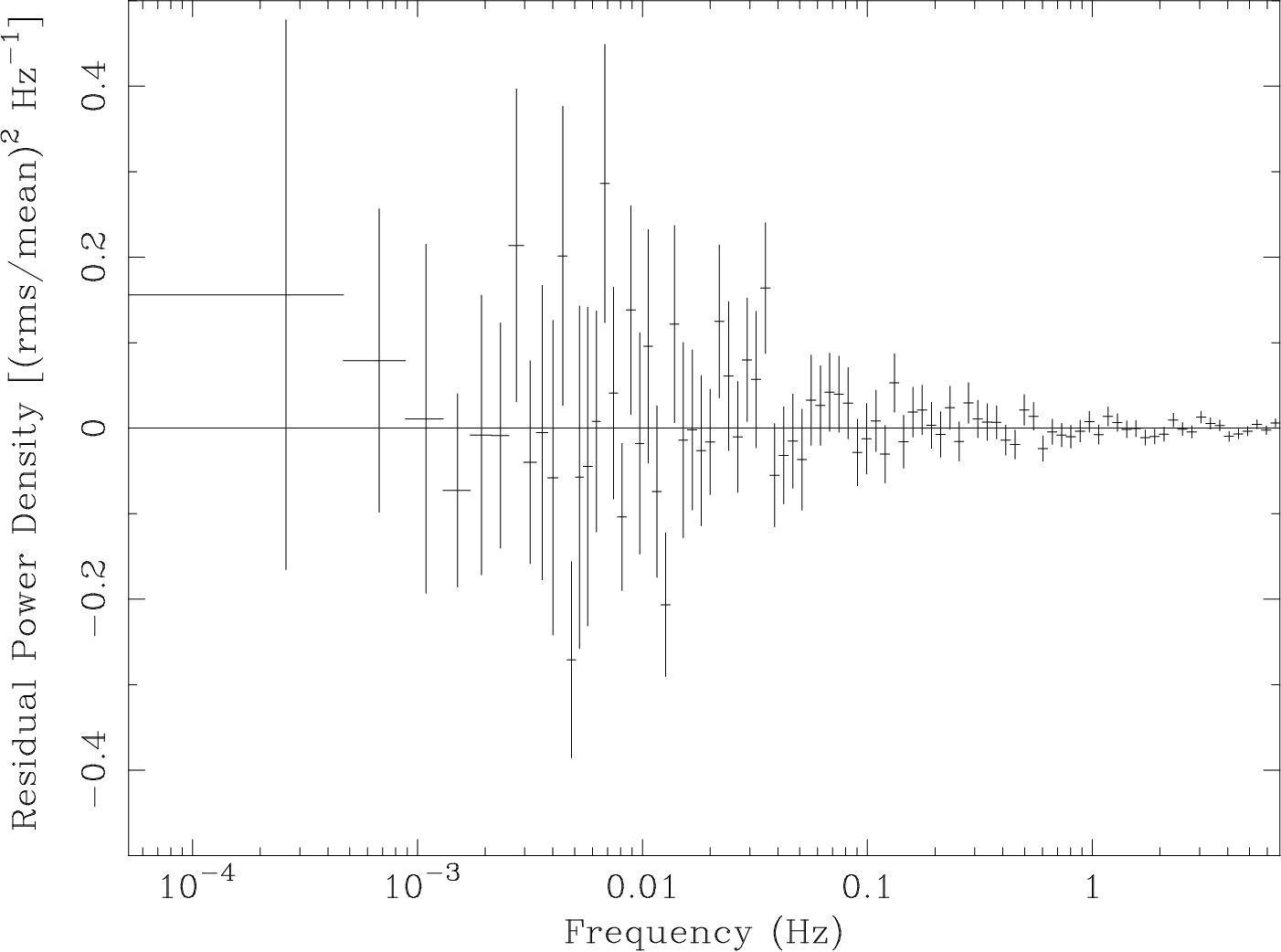}
\caption{Left panel: EPIC X-ray spectra of Ho II X-1.  The pn data
points are displayed in black, and the MOS in grey.  We show the
modelled contributions of the accretion disc (dotted lines) and corona
(solid lines) separately in the top part of the panel, and show the
model fit residuals in the lower part.  Right panel: the residual
power density after fitting the PSD of Ho II with a constant.  This
demonstrates the lack of variability power in the source.}
\label{fig2}
\end{figure*}

\subsection{Spectra}

Though more than 60\% of this observation was lost to space weather,
we were still able to extract the first reasonable signal-to-noise RGS
spectrum of an ULX.  This showed a smooth continuum shape, modelled
simply as an absorbed power-law continuum ($N_{\rm H} \sim 2 \times
10^{21}$ cm$^{-2}, \Gamma \sim 2.6$) with the exception of an excess
of counts slightly above 0.5 keV (see Fig. 1 of \citealt{GRRU05}).
This could be fit by an O VII triplet, but a much better solution was
found by allowing the abundance of the absorbing material to drop to
$\sim 0.6$ of the solar value.

Interestingly, this result strongly affects the EPIC data modelling.
In particular, using a 0.6-solar abundance {\tt TBABS} model in
\textsc{xspec} (and interstellar abundances set to the values of
\citealt{WAM00}) greatly reduces the size of the apparent soft excess,
and so the mass of the BH estimated from the IMBH model is reduced to
$\sim 33\%$ of its value assuming a solar abundance absorber.
However, the best fit to the data was again found to be the physically
self-consistent accretion disc + corona model, with $kT_{\rm in} \sim
0.2$ keV and $\tau \sim 4 - 9$ (i.e. a cool disc and optically thick
corona).  The EPIC spectral data, with this best-fit model, are shown
in Fig.~\ref{fig2}.  Note there is no detection of an Fe K$\alpha$
line in this dataset; formally, we place a 90\% upper limit of 25 eV
on the equivalent width of a narrow 6.4 keV Fe K$\alpha$ feature.

\subsection{Timing}

The EPIC timing data showed Ho II X-1 to be remarkably invariant
during the observation.  A power spectral density (PSD) analysis was
performed, finding that no power was evident (above the Poisson noise
level) in the $\sim 10^{-4} - 6$ Hz range (see Fig.~\ref{fig2}).  This
likely rules out Ho II X-1 being in a high/soft state, as a comparison
to Galactic black holes and AGN shows they generally have a red noise
spectrum with measureable RMS variability in this state\footnote{We
note that dilution by photons from the accretion disc should not be a
problem for Ho II X-1 if it contains an IMBH, as the 0.3--6\,keV band
we derive the PSD from is dominated by the putative coronal component.
However, it has been brought to our attention that there may be a
dimunition of intrinsic variability in the coronae of high/soft state
sources at the highest luminosities (J. Homan, {\it priv. comm.\/}),
which merits further quantification.}.  It does not rule out a state
with a band-limited PSD, such as occurs in the low/hard or some very
high states.  However, the strong limits placed by the non-detection
of power in the observed frequency interval implies any power must be
present at higher frequencies.  Assuming that BH timing properties
scale linearly with mass (e.g. \citealt{UMP02}), we can place an upper
limit on the mass of Ho II X-1 of 100 M$_{\odot}$ if it is in the low
or certain very high states.  Encouragingly, GRS 1915+105 shows very
similar variability characteristics in its ``$\chi$-class'' of
behaviour, which is thought to occur whilst it is in the very high
state.

\section{Radical solutions}

Whilst it is evident from this work that optically-thick coronae may
be common in ULXs, what is their origin?  One possible explanation is
offered by the model of \cite{Zhang00}, that explains accretion discs
as a two-layer system, with a cool ($0.2 - 0.5$ keV) interior seeding
a warm, optically thick ($1 - 1.5$ keV, $\tau \sim 10$) Comptonising
upper disc layer.  This model therefore offers an explanation for both
spectral components seen in our modelling.  Crucially, it has also
successfully been used to describe the X-ray spectrum of GRS 1915+105.
We therefore speculate that our spectral modelling, considered along
with the mass limit from Ho II X-1, argues that Ho II X-1 and many
other ULXs may be analogues of GRS 1915+105.  Whilst ULXs may still
possess larger BH masses than GRS1915+105 (i.e. 20 - 100 M$_{\odot}$,
so still technically IMBHs), we suggest they are operating in similar
accretion states, with a key link being that the accretion is
persistently at around the Eddington limit.

Obviously, we cannot rule out cool discs being the signature of a
larger, $\sim 1000$ M$_{\odot}$ IMBH.  Indeed, a weakness of the above
model is how we manage to see photons from the inner disc layer
through an optically thick exterior.  However, we note there are also
weaknesses with a literal interpretation of cool disc parameters, most
notably in the case of PG quasars.  Similar soft spectral components -
also modelled as cool discs - are seen in PG quasars, where their
temperature has been shown to be completely independent of the mass of
the BH (\citealt{GD04}).  This implies a radically different origin
for the soft excesses in PG quasars - and, by extension, ULXs - such
as in an outflow (as suggested in our dual-thermal model), or perhaps
even as atomic features on the accretion disc spectrum.

\section{Concluding comments}

If ULXs do contain $\sim 1000$ M$_{\odot}$ IMBHs, as suggested by
empirical modelling of their spectra, then one might reasonably assume
from their luminosities that they are accreting at $\sim 10\%$ of the
Eddington rate, and so should behave like scaled-up versions of BHXRBs
in the high-state.  The detection of spectral curvature in the
2--10\,keV range in the majority of our sample of ULXs, that can be
physically modelled as a cool, optically-thick corona, argues that
something different is happening in many ULXs.  One explanation could
be that these ULXs are operating in a fashion similar to GRS 1915+105,
as suggested by the timing results for Ho II X-1, in which case the
ULXs might harbour black holes not much more massive than in Galactic
BHXRBs.  This would bring X-ray spectral results more into line with
other arguments on ULXs that suggest that the majority of sources do
not contain IMBHs.  For example, \cite{King04} argues that the sheer
numbers of ULXs in the most extreme star-forming environments makes it
very unlikely that they can all contain IMBHs, and the modelling of
\cite{RPP05} demonstrates many ordinary high-mass BHXRBs (that one would
expect to find in these environments) have a mass transfer rate that
is substantially super-Eddington, providing an adequate reservoir of
fuel for ULXs.  Not least, it would agree with observations of
accreting sources in our own Galaxy that show many - most notably GRS
1915+105 - do experience episodes of radiating at a super-Eddington
level (cf.. \citealt{MR03}).

However, we certainly cannot exclude ULXs from containing $\sim 1000$
M$_{\odot}$ IMBHs at the current point in time, and indeed two of the
sources in our sample do appear consistent with this interpretation.
Ultimately it may prove that the only measurement that can resolve the
debate on the underlying nature of ULXs will be a dynamical mass limit
for the black hole from its orbital motion, as is the case for
Galactic BHXRBs.  Until then, more and deeper X-ray observations are
required to advance our knowledge of these extraordinary X-ray
sources.

\section*{Acknowledgments}

TPR, AMS and MRG gratefully acknowledge funding from PPARC.  This work
is based on observations obtained with {\it XMM-Newton\/}, an ESA
science mission with instruments and contributions directly funded by
ESA member states and NASA.

\end{document}